\documentclass[a4paper,fleqn,usenatbib]{mnras}
\usepackage{ulem}
\usepackage[utf8]{inputenc}

\newcommand{\II}{\small{II}\normalsize}

\usepackage{newtxtext,newtxmath}

\usepackage[T1]{fontenc}
\usepackage{subfigure}
\usepackage{ae,aecompl}
\usepackage{siunitx}
\usepackage{soul}

\usepackage{graphicx}	
\usepackage{amsmath}	
\usepackage{amssymb}	
\usepackage{enumerate}
\usepackage{natbib}
\usepackage{adjustbox}
\usepackage{helvet,soul}
\usepackage[font=small]{caption}

\title[SZ Pictoris]{The active RS CVn-type system SZ Pictoris}

\author[C. I. Mart\'inez et al.]{
C. I. Mart\'inez$^{1,2}$\thanks{Based on data obtained at Complejo Astron\'omico El Leoncito, operated under agreement between the Consejo Nacional de Investigaciones Cient\'ificas y T\'ecnicas de la Rep\'ublica Argentina and the National Universities of La Plata, C\'ordoba and San Juan.},
J. F. Gonz\'alez$^{2}$, A. Buccino$^{3,4}$,
R. Iba\~nez Bustos$^{3}$ and
\newauthor P. J. D. Mauas$^{3,4}$
\\
$^{1}$Observatorio Astron\'omico F\'elix Aguilar, Universidad Nacional de San Juan, Av. Benavides 8175 oeste, San Juan, Argentina.\\
$^{2}$Instituto de Ciencias Astron\'omicas de la Tierra y el Espacio (CONICET-UNSJ), Av. España 1512 sur, San Juan, Argentina.\\
$^{3}$Instituto de Astronom\'\i a y F\'\i sica del Espacio (CONICET-UBA), C.C. 67 Sucursal 28, C1428EHA-Buenos Aires, Argentina. \\
$^{4}$Departamento de F\'\i sica. Facultad de Ciencias Exactas y Naturales - Universidad de Buenos Aires, Buenos, Argentina.\\
}

\date{Accepted 24 October 2019}

\pubyear{2019}

\begin{document}
\label{firstpage}
\pagerange{\pageref{firstpage}--\pageref{lastpage}}
\maketitle

\begin{abstract}
We study the short and long-term variability of the spectroscopic binary SZ Pictoris, a southern RS CVn-type system. We used mid-resolution \textit{echelle} spectra obtained at CASLEO spanning 18 years, and photometric data from the ASAS database (\textit{V}-band) and from the Optical Robotic Observatory (\textit{BVRI}-bands) for similar time lapses. 
We separated the composite spectra into those corresponding to both components, and we were able to determine accurate orbital parameters, in particular an orbital period of 4.95 days. We also observed a photometric modulation with half the orbital period, due to the ellipticity of the stars.
We also found cyclic activity with a period of \(\sim\) 2030 days, both in the photometry and in the Ca II flux of the secondary star of the system.
\end{abstract}

\begin{keywords}
stars: individual: SZ Pictoris - stars: activity - stars: fundamental parameters.
\end{keywords}



\section{Introduction}

RS CVn stars are close detached binaries formed by a giant or subgiant star of spectral types G to K as the more massive primary component and a subgiant or dwarf star of spectral types G to M as the secondary. Since these stars are fast rotators tidaly synchronized, they show higher levels of activity than single stars with the same characteristics. In particular, these systems typically exhibit strong Ca \II\-, H\(\alpha\), X-ray, and microwave emissions \citep{Hall1976}. At optical wavelengths, the most prominent feature of the RS CVn stars is its periodic photometric variability, which is thought to be the consequence of rotational modulation by large dark starspots (see for example \citealt{Roettenbacher2016}). On the other hand, many of these stars show long-term chromospheric activity variations and activity cycles \citep{Buccino2009}.

SZ Pictoris is a southern star reported as a double-lined spectroscopic binary and photometric variable of the RS CVn type by \citet{Andersen1980}. They found that both components have similar spectral types and exhibit  Ca \II\- H and K line-core emission. The apparent magnitude is $V = 7.893$ mag, and the spectral line ratios correspond to a luminosity ratio of 0.7 - 0.8 \citep{Andersen1980}. The effective temperature $T_\mathrm{eff} = 5310$ K and the lithium abundance $\log n(Li) < 1.8$ were determined by \citet{Pallavicini1992} using high-resolution ($R = 50.000$) and high S/N (\(\geq\) 100) spectra.

\citet{Cutispoto1998a} determined a spectral type K0/1 IV + G5 IV for the system. However, he later found that higher luminosities were required to fit the \textit{Hipparcos} distance (172-224 pc, \citealt{Perryman1997}) and he adopted a K0 IV/III + G3 IV/III \citep{Cutispoto1998b}. The more precise distance obtained by the Gaia mission ($d = 188.4 \pm 0.9$ pc, \citealt{gaia2018}) confirms this luminosity-class assignment.

\citet{Rocha-Pinto1998} obtained an activity level for the system of $\log R^\prime_\mathrm{HK} = - 4.05$ and a difference between photometric and spectroscopic metallicity of $\Delta[Fe/H] = -0.444$. Later, \citet{Rocha-Pinto2000} suggested that the low photometric metallicity can be explained as an effect of the reddening, since at SZ Pic's distance we should expect that the object is mildly to strongly reddened resulting in a photometric metallicity that would seem lower than it really is. They calculated the reddening in the region of SZ Pic using $\beta$ index from \citep{hauck1998}, the intrinsic colour calibration of \citep{Schuster1989}, and the \textit{Hipparcos} distance 194.9 pc, obtaining $E(b-y) = 0.064$.

Using a method based on the concept of common chromospheres (\textit{roundchroms}), \citet{Gurzadyan1997} determined an intercomponent distance of 14.9 R\textsubscript{\(\odot\)}, as well as an electron concentration in the \textit{roundchroms} of $n_\mathrm{e} = \num{7e10}$ cm\textsuperscript{-3}. This method supposed a roundchrom enveloping both components of the binary system without coming into contact with their photospheres. It is based on the identification of the outer boundary of a roundchrom with an equipotential zero velocity surface at some value of the Jacobi constant $C$, at which the formation of a narrow corridor near the central Lagrangian point $L_2$ will be ensured for the transit of gaseous matter from the main component of the system in the direction of the secondary component. The central concept of the determination of the radius of the main component of the system is based on the assumption that the observed Mg \II\- emission ($\lambda = 2800$ \AA) is entirely of roundchrom origin \citep{Gurzadyan97a}.

In 1999 we started the \textit{HK\(\alpha\) Project} to study the long-term chromospheric activity in late-type stars using medium-resolution spectra obtained at the Complejo Astron\'omico El Leoncito (CASLEO), San Juan, Argentina. In the framework of this project, for SZ Pic we measured the activity index $S_\mathrm{MW} = 0.54$ \citep{Cincunegui2007b} contrasting with the earlier observation by \citet{Henry1996}, who obtained $S_\mathrm{MW} = 0.975$ . In this work, we present spectra of SZ Pic spanning more than eighteen years, together with two photometric data sets obtained by the All Sky Automated Survey (ASAS, \citealt{Pojmanski2002}) and our own observations.
In Section 2 we present an overview of our spectroscopic and photometric observations and the reduction methods employed. In Section 3 we describe the method used to study the temporal evolution of the data. In Section 4 we describe the results and, finally, we discuss the results and possible future work in Section 5.

\section{Observations and Data Reduction}

Since 2000, we have been continuously observing SZ Pic as part of the ongoing HK\(\alpha\) Project. This data set allows us to have a unique long time-series of Mount Wilson indexes, which is the most extended activity indicator used to detect stellar activity cycles (\citealt{Baliunas1995}; \citealt{Cincunegui2007}; \citealt{Metcalfe2013}; \citealt{Flores2017}). The observations were obtained with the REOSC spectrograph\footnote{http://www.casleo.gov.ar/instrumental/js-reosc.php} ($R= \lambda / \Delta\lambda \approx 13.000$), mounted on the 2.15 m JS telescope at the CASLEO. This equipment provides us with mid-resolution echelle spectra covering a wavelength range between 3800 and 6900 \AA, which allows us to simultaneously study the effect of chromospheric activity in the whole optical spectrum. These echelle spectra were optimally extracted and flux calibrated as explained in \citet{Cincunegui2004}. The resulting spectral resolution is high enough to clearly resolve the double features in the spectra of SZ Pic at appropriate phases.

 Table \ref{dataspec} lists the spectral observation logs of SZ Pic. The first and third column show the date (month and year as MMYY) of the observations and the second and fourth column list $xJD = HJD - 2450000$, where HJD is the heliocentric Julian date at the beginning of the observation. There is a total of 26 individual observations distributed from August 2000 to November 2018. 

\begin{table}
\centering
\caption{Logs of the CASLEO Observations of SZ Pic. Column 1 and 3: date of each observing run (MMYY). Column 2 and 4: Heliocentric Julian date xJD = HJD - 2,450,000.}
\begin{tabular}{cccccc}
\hline\noalign{\smallskip}
Label	&	xJD &	&	& Label	&	xJD\\
\hline\noalign{\smallskip}
0800 & 1770 &	&	& 1206 & 4081 \\
0301 & 1973 &	&	& 1208 & 4821 \\
0302 & 2364 &	&	& 0309 & 4904 \\
0802 & 2520 &	&	& 0310 & 5264 \\
1102 & 2601 &	&	& 0311 & 5638 \\ 
0303 & 2715 &	&	& 1212 & 6282 \\
0903 & 2898 &	&	& 1212 & 6285 \\
1203 & 2981 &	&	& 0313 & 6356 \\
0304 & 3074 &	&	& 1013 & 6592 \\
0904 & 3277 &	&	& 0314 & 6736 \\
1104 & 3336 &	&	& 1214 & 7007 \\
0305 & 3449 &	&	& 1118 & 8436 \\
1105 & 3700 &	&	& 1118 & 8437 \\
\hline 
\end{tabular}
\label{dataspec}
\end{table}

We also carried out photometric observations of SZ Pic using the four 16'' MEADE telescopes that form the Optic Robotic Observatory (ORO) installed at the Observatorio Felix Aguilar (OAFA), also located in San Juan, Argentina. We observed SZ Pic for 4-5 hours per night in non-consecutive nights to have a near uniform sampling throughout the expected orbital phase. During each observing night, we obtained science images every 5 minutes approximately. 
In total, we obtained 797 good images between March 2018 and January 2019. We estimated the formal error for each observation as the sum in quadrature of the errors of the fluxes of the target and the reference star. The maximum errors were near 17 mmag.

\begin{figure}
  \centering
  \includegraphics[width=\columnwidth]{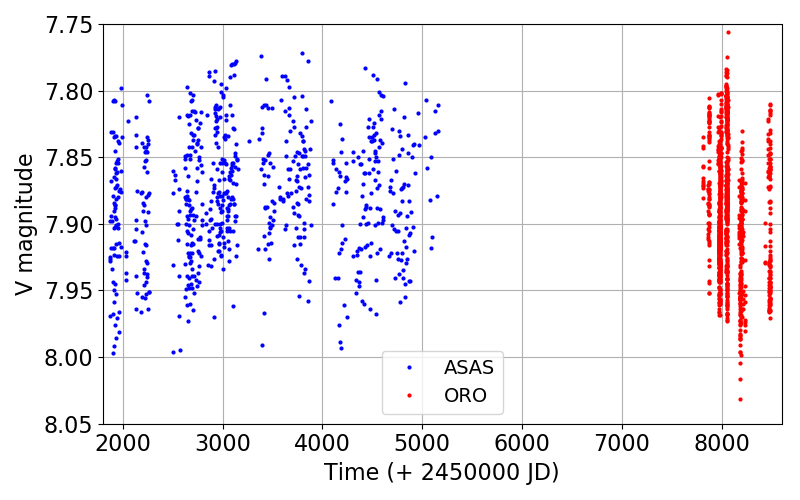}
  \caption{$V$ magnitude of SZ Pic as a function of time (ASAS in blue; ORO in red).}
  \label{asas_mate}
 \end{figure}

In addition, we also employed photometric observations provided by the ASAS-3 database. 
We only used observations qualified as A or B in the database ("best" or "mean" quality according to the ASAS definition). The final time series consists of 734 points, between November 2000 and November 2009, with typical errors of around 31 mmag. In Fig. \ref{asas_mate} we plot the $V$ magnitude of SZ Pic as a function of time, obtained with ASAS-3 and ORO. Similarly to our previous works \citep{Diaz2007,Buccino2011,Buccino2014,Ibanez2019}, we chose the mean $V$ magnitude of each observing season as a proxy for magnetic activity .

\section{Orbital parameters of the system}

\begin{figure}
  \centering
  \includegraphics[width=\columnwidth]{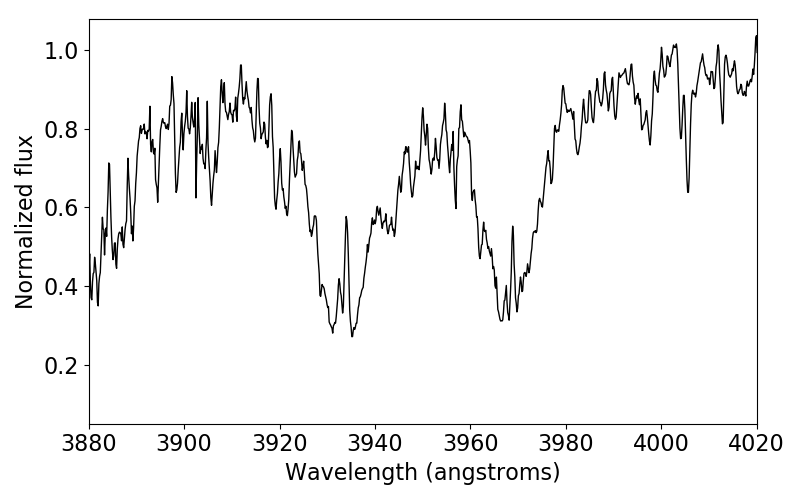}
  \caption{Example of an original composite spectrum in the Ca II region.}
  \label{specori}
\end{figure}

In Fig. \ref{specori} we show an example of the original spectra, corresponding to March 2001, in the Ca \II\- region. The double emission peaks can be easily noted. We processed these original spectra using the iterative method developed by \citet{Gonzalez2006}. This method allows us to separate the individual spectra of each component, and to compute their radial velocities (RVs). In each iteration, the spectrum computed for one component is removed from the observed one. The resulting single-lined spectrum is then used to measure the RV of the remaining component and to compute its spectrum by combining them appropriately.

\begin{figure}
  \centering
  \label{comp}
\subfigure[\label{compa}]{\includegraphics[width=\columnwidth]{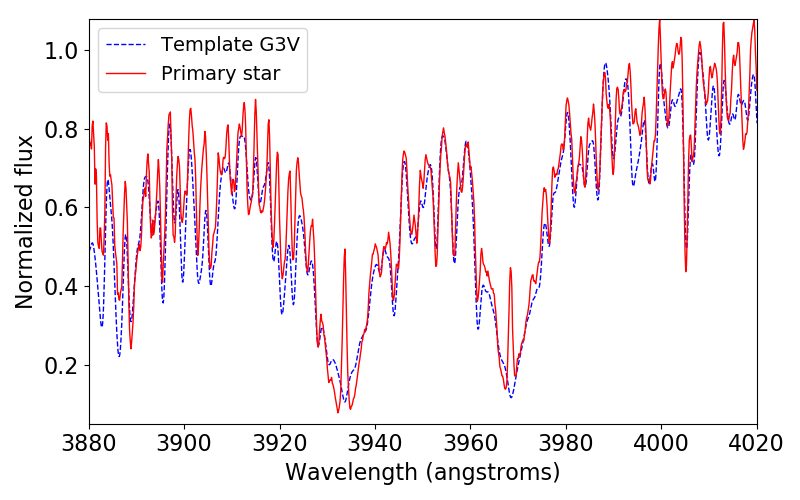}}
\subfigure[\label{compb}]{\includegraphics[width=\columnwidth]{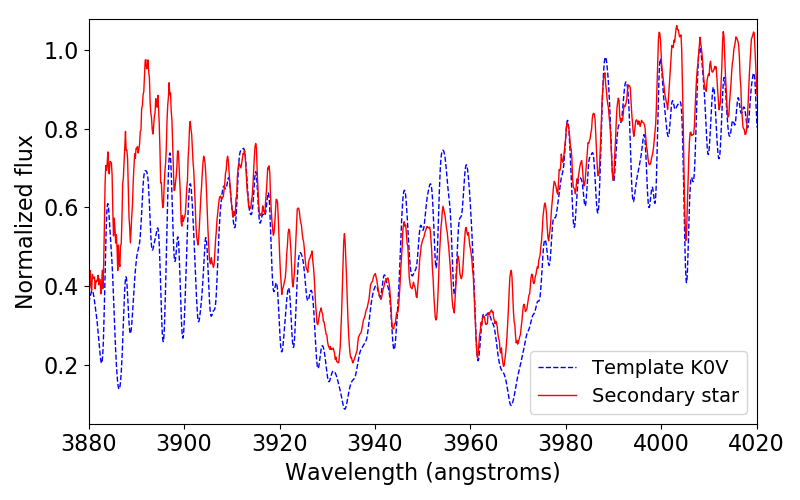}}
  \caption{Mean spectra in the Ca II region of each component (red solid line) compared with the corresponding templates (blue dashed line).}
  \label{comp_spec}
\end{figure}

We employed the mid-resolution \textit{G\"ottingen Spectra Libraries by Phoenix}\footnote{http://phoenix.astro.physik.uni-goettingen.de} \citep{Husser2013} to select the template for the cross-correlations, according to the spectral classification by \citet{Cutispoto1998b} mentioned in Section 1. In all cases, for the RV measurements we excluded the emission lines.  A mean spectrum of very high S/N (\(\geq\)100) was obtained for each component. In Fig. \ref{comp_spec} we show both spectra compared with the templates in the Ca \II\- H and K region. 

The Lomb–Scargle (LS) periodogram \citep{Horne1986} has been extensively employed to search for stellar activity cycles (e.g. \citealt{Baliunas1995}, \citealt{Metcalfe2013}, \citealt{Flores2017}).  Recently, \citet{Zechmeister2009} proposed a modification, the Generalized Lomb–Scargle (GLS) periodogram, which has certain advantages in comparison to the classic LS periodogram: It takes into account a varying zero point, it does not require a bootstrap or Monte Carlo algorithms to compute the significance of a signal, reducing the computational cost, and it is less susceptible to aliasing than the LS periodogram. 

\begin{figure}
  \centering
  \includegraphics[width=\columnwidth]{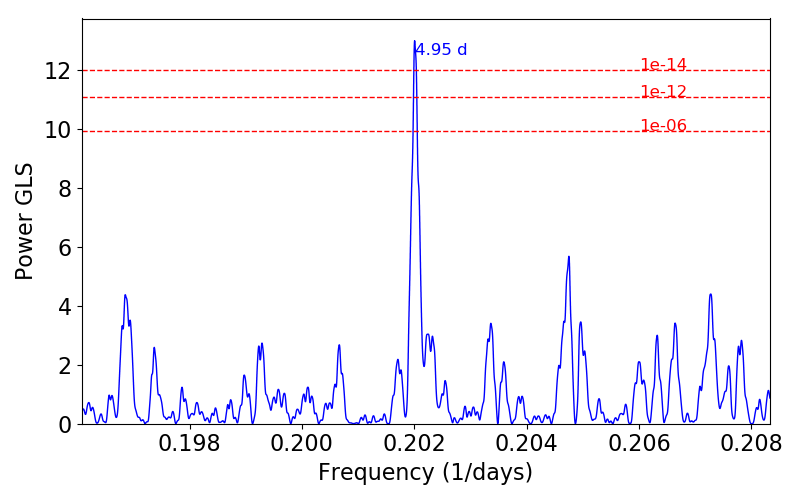}
  \caption{GLS periodogram of the differential radial velocities between the primary and secondary components. The most significant peak is indicated. The dashed lines indicate the false alarm probability levels.}
  \label{period_orbital}
\end{figure}

We calculated the GLS periodogram to the RV difference between the components, to eliminate the possible systematic errors or eventual variations of the systemic velocity (see Fig. \ref{period_orbital}). The amplitude of the periodogram at each frequency point is identical to the value that would be obtained estimating the harmonic content of a data set, at a given frequency, by linear least-squares fitting to a harmonic function of time \citep{Press1992}. 

To estimate the false alarm probabilities (FAP) of each peak of the periodogram we use:
\begin{equation}
FAP = 1- \left[ 1 - \left( 1 - \frac{2P}{N-1} \right)^{\frac{N-3}{2}} \right] ^M
\label{intcuad}
\end{equation}

where $N$ is the size of the dataset, $M$ is the number of independent frequencies and $P$ is the power of the period detected in the GLS periodogram \citep{Zechmeister2009}. We normalize the periodogram assuming Gaussian noise.

\begin{figure}
  \centering
   \subfigure[\label{vdiff}]{\includegraphics[width=\columnwidth]{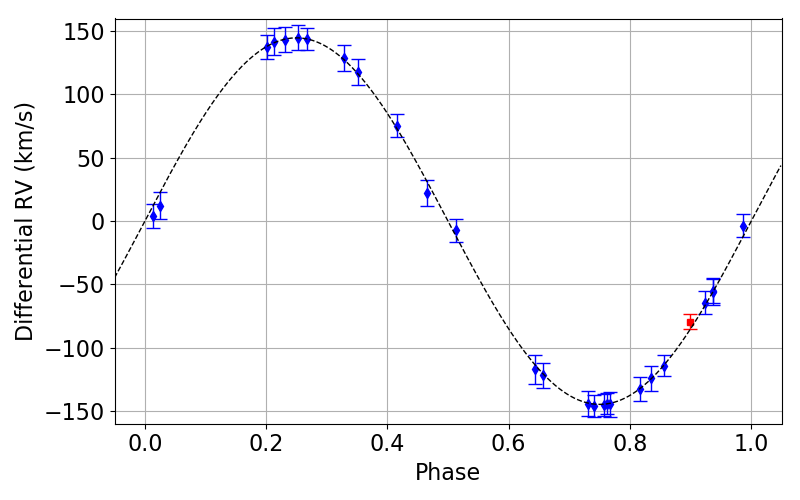}}
  \subfigure[\label{vra_vrb}]{\includegraphics[width=\columnwidth]{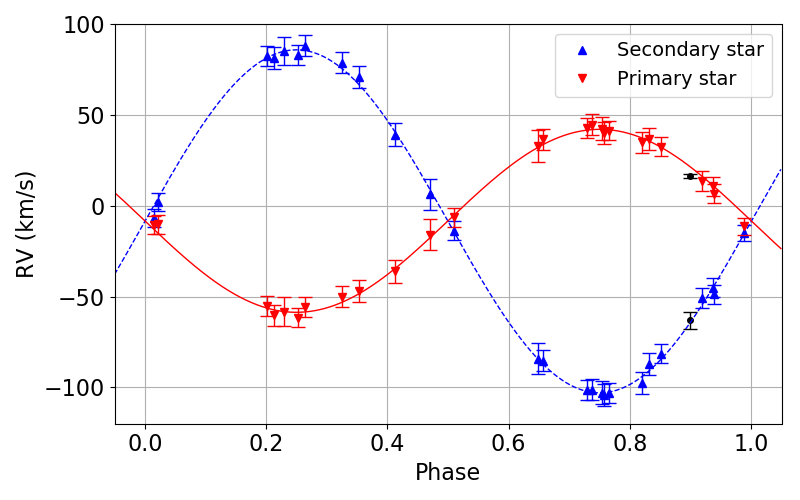}}
  \caption{RV measurements of each component; the black points correspond to individual RVs determined by \citet{Andersen1980}, $16.2 \pm 1.2$ km/s and $-63.1 \pm 4.5$ km/s in JD = 2,443,789.8801 (October 1978). In both figures, the dashed lines are the best fits to the data.}
  \label{phase_vdiff}
\end{figure}

We obtained an orbital period $P = 4.950320 \pm 0.000014$ days, with a FAP smaller than $10^{-14}$. Using this period, we fit a Keplerian orbit to the RVs of the system (see Fig. \ref{phase_vdiff}). The orbital parameters are listed in Table \ref{orbital_param}.

By comparing the mean spectra of the two components with synthetic spectra from the \textit{G\"ottingen Spectra Library} of \citet{2013A&A...553A...6H} based on \textit{Phoenix} atmosphere models, we estimated the effective temperatures and flux ratio between the components, obtaining: $T_\mathrm{eff}(A) = 5700 \pm 300$ K, $T_\mathrm{eff}(B)= 5400 \pm 300$ K, and $f_A/f_B = 2.24 \pm 0.33$. Additionally, we computed the absolute magnitude using the apparent magnitude and the Gaia distance ($d = 188.4 \pm 0.9$ pc), obtaining $M_\mathrm{V} = 1.46 \pm 0.10$ mag for the whole system. In these calculations the interstellar extinction ($A_\mathrm{V}=0.074$ mag) was calculated from  the \citet{Schlegel1998} maps and the distance, applying the same procedure as \citet{bilir2008}.

Then, using the flux ratio and appropriate bolometric corrections \citep{flower1996} we estimate the absolute bolometric magnitude for each component: $2.64 \pm 0.16$ mag for primary star and $1.70 \pm 0.12$ for secondary star. Finally, we derive the stellar radius of the components from their luminosities and temperatures: $2.7 \pm 0.4$ $R_\odot$ for primary and $4.6 \pm 0.5$ $R_\odot$ for secondary. We found no evidence of eclipses in the photometry, therefore, we can set an upper limit for the inclination angle. From the spectroscopic parameter $a\sin{i} = 14.14$ $R_\odot$ and the estimated radii, it can be inferred that the inclination angle $i$ is lower than 63 deg. In turn, this imposes lower limits for the stellar masses: $M_1 > 1.54$ and $M_2 > 0.82$ $M_\odot$. In Table \ref{fisicos} we summarize the intrinsic properties of the components of SZ Pictoris.

\begin{table}
\centering
\caption{Orbital parameters for SZ Pic.} 
\begin{tabular}{lr}
\hline\hline\noalign{\smallskip}
Conjunction epoch [HJD] & 2,451,1111.1903 \(\pm\) 0.0059 \\
Period [days] & 4.950320 \(\pm\) 0.000014 \\
\textit{a} sin \textit{i} [\textit{R}\textsubscript{\(\odot\)}] & 14.14 \(\pm\) 0.14 \\
Systemic velocity [km\,s$^{-1}$] & -8.47 \(\pm\) 0.21 \\
Primary RV amplitude [km\,s$^{-1}$] & 50.36 \(\pm\) 0.78 \\
Secondary RV amplitude [km\,s$^{-1}$] & 94.40 \(\pm\) 0.69 \\
Orbital excentricity & 0.016 \(\pm\) 0.007 \\
\textit{M}\textsubscript{1} sin\textsuperscript{3} \textit{i} [\textit{M}\textsubscript{\(\odot\)}] & 1.014 \(\pm\) 0.015 \\
\textit{M}\textsubscript{2} sin\textsuperscript{3} \textit{i} [\textit{M}\textsubscript{\(\odot\)}] & 0.54 \(\pm\) 0.011 \\
\hline
\end{tabular}
\label{orbital_param}
\end{table}

\begin{table}
\centering
\caption{Intrinsic properties of SZ Pic.} 
\begin{tabular}{lcc}
\hline\hline\noalign{\smallskip}
	&	Primary Star	&	Secondary Star	\\
\noalign{\smallskip}
$T_\mathrm{eff}$ [\textit{K}]	&	5700  \(\pm\) 300	&	5400  \(\pm\) 300	\\
\textit{L} [\textit{L}\textsubscript{\(\odot\)}]	&	6.5 \(\pm\) 1.0	&	15.4 \(\pm\) 1.8	\\
\textit{R} [\textit{R}\textsubscript{\(\odot\)}]	&	2.7 \(\pm\) 0.4	&	4,6 \(\pm\) 0.5	\\
\textit{M} [\textit{M}\textsubscript{\(\odot\)}]	&	> 1.54	&	> 0.82	\\
\hline
\end{tabular}
\label{fisicos}
\end{table}

\begin{figure}
  \centering
  \includegraphics[width=\columnwidth]{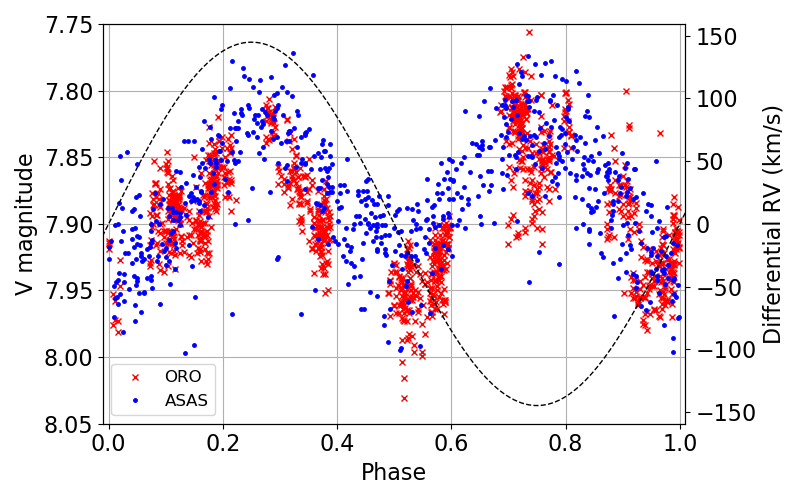}
  \caption{Photometric variation for ASAS (blue) and ORO (red) together with the fit to the rotation (black dashed line).}
  \label{vel_phot}
\end{figure}

To search for rotational modulations in the photometric data, we applied the GLS periodogram to the photometric observations described in Sect. 2. In Table \ref{comparacion_periodos} we summarize the results obtained for the $B$, $V$, $R$, and $I$ bands data observed with ORO, and the $V$ observations from ASAS. We see that the photometric periods obtained agree very well, and are exactly half the orbital period, with errors smaller than 0.03\% and very low FAP levels. The results are shown in Fig. \ref{vel_phot}, where we show the photometry phased to the orbital period together with the Keplerian orbit. It can be seen that the system is brighter when the velocity is maxima, {\it i.e.} at quadrature. This result agrees with the hypothesis formulated by \citet{Drake1989}, that the photometric period for these synchronized systems should be half the orbital period; and most of the variability is caused by an elipticity effect on the shape of the stars \citep{Cutispoto1998b}. 

\citet{Bell1983} estimated a photometric period of $P_\mathrm{phot} = 2.441$ days with an amplitude of the light curve of about 0.15 mag, using the data set of \citet{Andersen1980}. Later, \citet{Strassmeier1993} found an orbital period $P_\mathrm{orb} = 4.92$ d, and \citet{Cutispoto1995} obtained a photometric period $P_\mathrm{phot} = 4.905$ d. All these values agree very well with our results.

\begin{table*}
\centering
\caption{GLS analysis of the different photometry data sets.}
\begin{adjustbox}{max width=\textwidth}
\begin{tabular}{cccc}
\hline\hline\noalign{\smallskip}
Data	&	Photometric period	& Amplitude &	Magnitude \\
		&	(days)	& (days)	 &	(mag)     \\
\hline\noalign{\smallskip}
$B$-band (ORO)  & 2.4758 $\pm$ 0.0004  &  0.0501 $\pm$  0.0016  &  8.6590 $\pm$ 0.0011  \\
$V$-Band (ORO)  & 2.4755 $\pm$ 0.0002  &  0.0534 $\pm$  0.0011  &  7.8931 $\pm$ 0.0008  \\
$R$-band (ORO)  & 2.4756 $\pm$ 0.0003  &  0.0572 $\pm$  0.0022  &  7.7328 $\pm$ 0.0015  \\
$I$-band (ORO)  & 2.4759 $\pm$ 0.0004  &  0.0525 $\pm$  0.0017  &  7.4161 $\pm$ 0.0012  \\
ASAS-3 & 2.4752 $\pm$ 0.0001 &  0.0472 $\pm$  0.0018  &  7.8777 $\pm$ 0.0011  \\
\hline 
\end{tabular}
\end{adjustbox}
\centering
\label{comparacion_periodos}
\end{table*}

\section{Stellar Activity}

\begin{figure}
  \centering
  \subfigure[\label{long}]{\includegraphics[width=\columnwidth]{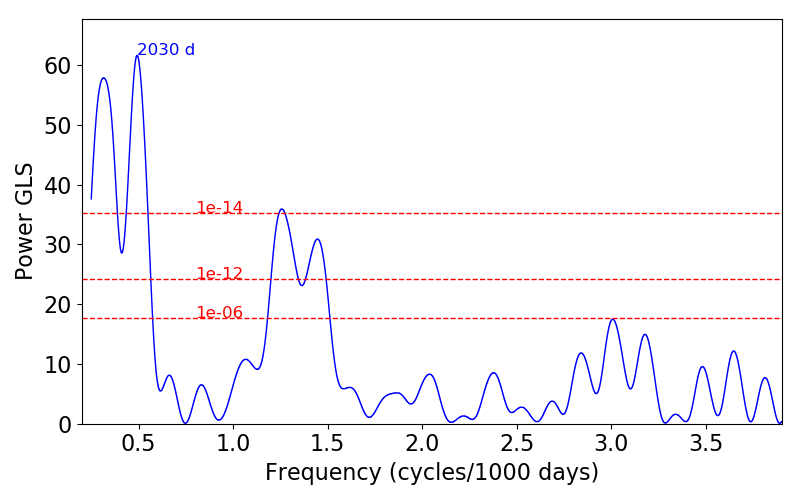}}
  \subfigure[\label{fase}]{\includegraphics[width=\columnwidth]{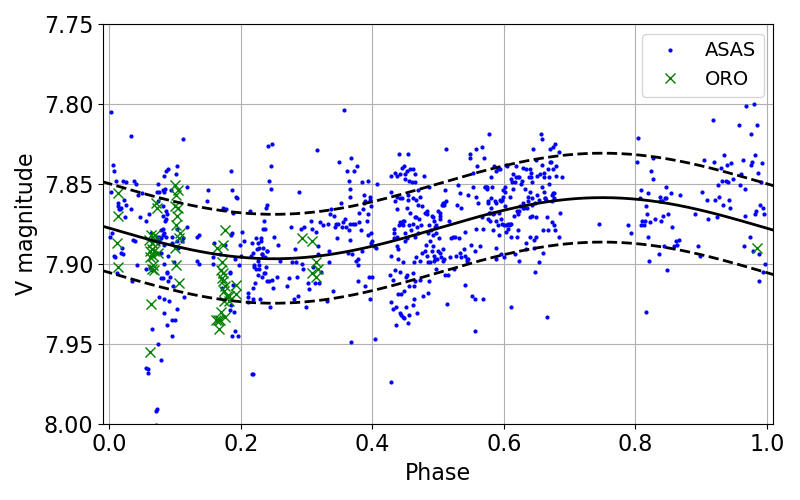}}
  \caption{(a) GLS periodogram of the complete photometry data set; the most significant peaks are indicated; the dashed lines indicated the FAP levels. (b) Phase curve for $P = 2030$ d; the ORO photometry is represented by the mean magnitude (green markers) and the dashed lines represent \(\pm\)\(\sigma\) deviations.}
  \label{long_per}
\end{figure}

To search for a long period, associated to a stellar activity cycle, we first removed from the complete photometry the photometric period $P = 2.47$ days found in Fig. \ref{period_orbital}, since it is too strong and masks any other variation. Then, we computed the GLS periodogram to the residuals. As can be seen in Fig. \ref{long_per}, there are two distinct peaks: the most significant peak corresponds to a period of $P = 2030 \pm 34$ days and the second peak corresponds to a period of $P = 3167 \pm 166$ days. For both peaks, the FAP is lower than \(\num{1e-14}\). However, the length of the ASAS dataset is of 3289 days. Therefore, we discard this second period as due to this effect. In Fig. \ref{fase} we show the photometric observations phased with the period of 2030 days.

\begin{figure}
  \centering
  \includegraphics[width=\columnwidth]{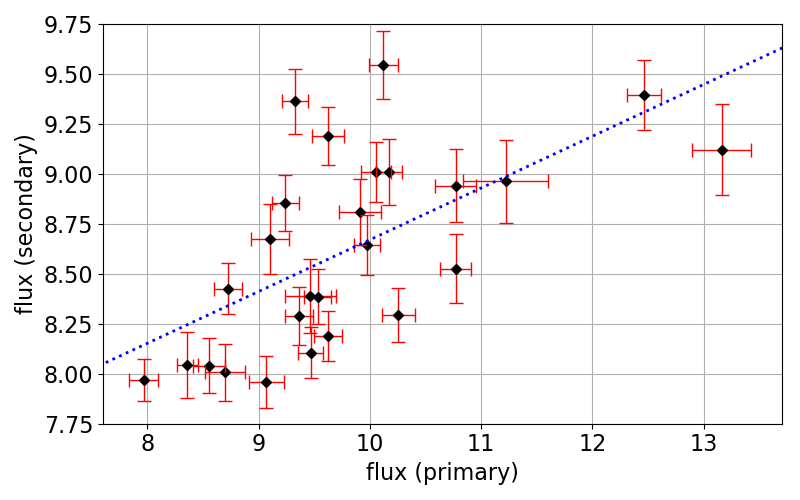}
  \caption{Comparison of the fluxes in the Ca II H and K lines for both components. We use the S/N ratio of each spectrum in the Ca II region to estimate the flux errors. Flux units are [\(\num{e-12}\) erg cm\textsuperscript{-2} \AA\textsuperscript{-1} s\textsuperscript{-1}].}
  \label{flux_a_b}
\end{figure}

To study the chromospheric activity of the system we use the flux in the Ca \II\- H and K lines. Since we have separated the spectra of both components, we should be able to distinguish whether the changes observed in the photometric measurements belong to one of the components or to both. The spectroscopic measurements also provide further information about the active regions present in the stellar surfaces. We carefully examined the separated spectra and discarded the spectra exhibiting signs of reduction problems or transient events, such as flares. To do this we compared the fluxes of the H and K lines in the individual spectra taken for each observation (see Sect. 2) and we discarded the pairs for which the difference were too large. 

The fluxes in the Ca \II\- H and K line-core emissions were integrated with a triangular profile of 1.09 \AA, as is usually done to compute the Mount Wilson \textit{S}-index. However, we did not normalize the flux by the continuum windows, since the continua is much weaker, and this would only introduce a large error factor. In Fig. \ref{flux_a_b} we plot the flux for both components of the system and the linear regression between them. It can be seen that both fluxes are fairly correlated, with a Pearson correlation coefficient $r = 0.62$. 

\begin{figure}
  \centering
  \subfigure[\label{per_flux}]{\includegraphics[width=0.9\linewidth]{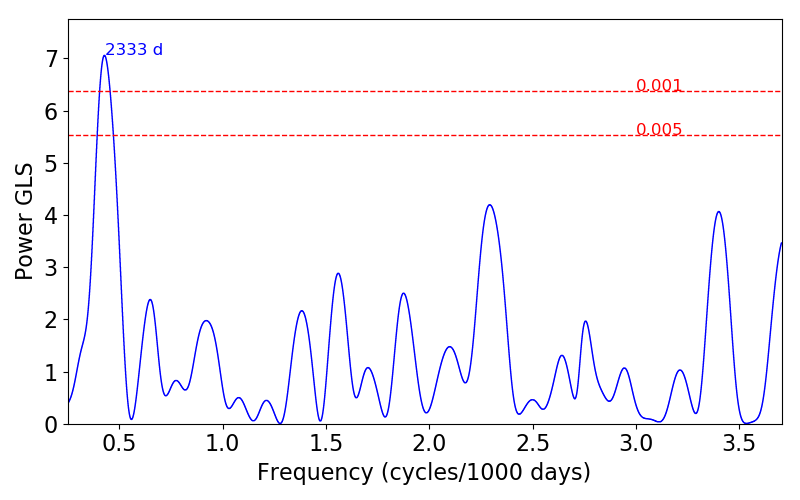}}
  \subfigure[\label{fase_flux}]{\includegraphics[width=\columnwidth]{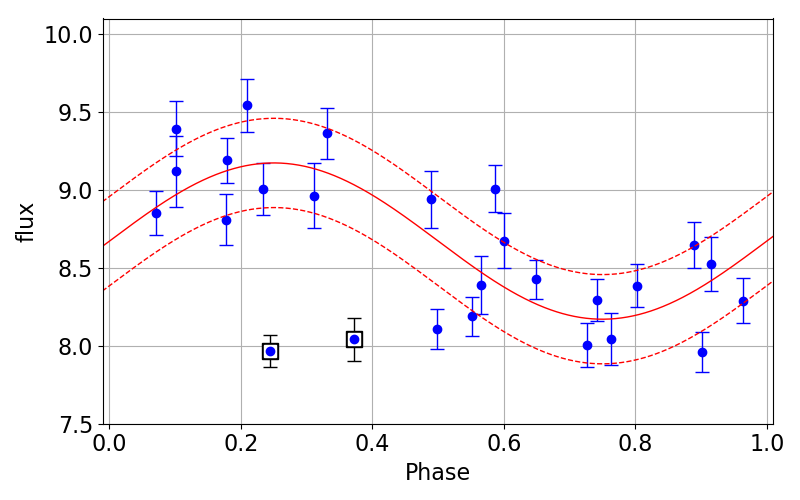}}
  \caption{(a) GLS periodogram of the Ca II H and K flux for the secondary component; the most significant peak is indicated; the dashed lines indicated the FAP levels. (b) Phase curve for the $P = 2333$ d (the dashed lines represent \(\pm\)\(\sigma\) deviations). We use the S/N ratio of each spectrum in the Ca II region to estimate the error fluxes. The square points are the outliers mentioned in the text. The units are [\(\num{e-12}\) erg cm\textsuperscript{-2} \AA\textsuperscript{-1} s\textsuperscript{-1}].}
  \label{figura9}
\end{figure}

We also computed the periodogram for the Ca \II\ flux of each component. In Fig. \ref{figura9} we show it for the secondary star. In this case a period $P = 2333 \pm 98$ with a FAP 0.0002 can be seen. There seem to be two outliers in the phase curve, which are responsible for most of the error. They were removed from the calculations. On the other hand, we found no evidence of a period for the primary component. 

From the separated CASLEO spectrum we computed the $\log R^\prime_\mathrm{HK}$ \citep{Noyes1984} indicator for each component. We obtained $\log R^\prime_\mathrm{HK} = -4.073$ for the secondary, the G-type star, and $\log R^\prime_\mathrm{HK} = -4.11$ for the primary, the K-type star. These values indicate that the secondary is the most active star of the system, and that a typical $\alpha\Omega$ dynamo could be responsible for the chromospheric cycle detected.

\begin{figure}
  \centering
  \includegraphics[width=\columnwidth]{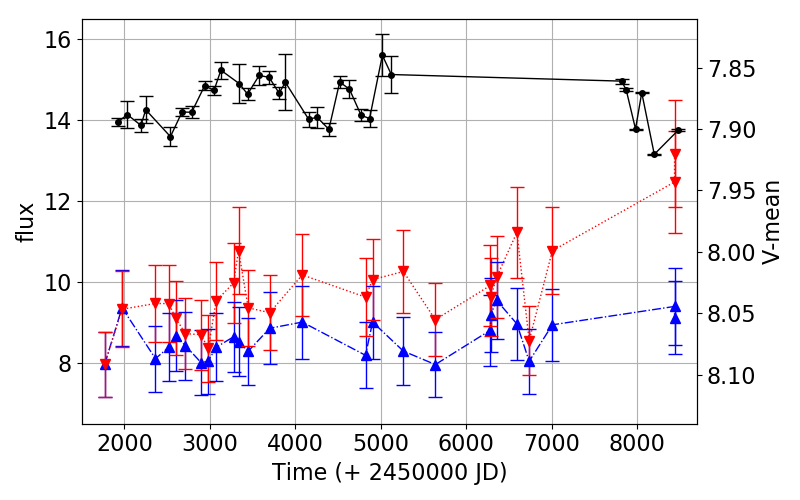}
  \caption{Mean magnitude (black solid lines), Ca II H and K flux of the primary component (red dotted line) and of the secondary component (blue dashed line) as a function of time. The error bars correspond to a 10\% error in the line fluxes. Flux units are [\(\num{e-12}\) ergs cm\textsuperscript{-2} \AA\textsuperscript{-1} s\textsuperscript{-1}].}
  \label{flux_temp}
\end{figure}

Finally, in Fig. \ref{flux_temp} we plot the flux as a function of time for both components, together with the mean magnitude of the system. Note that the maximum for the secondary component took place in 2013 (HJD = 2,456,355.5412).

\section{Discussion}

The aim of this paper is to study the short and long-term variability of the spectroscopic binary system SZ Pictoris. For this purpose, medium resolution \textit{echelle} spectra were collected at the CASLEO observatory, whereas photometric measurements in \textit{BVRI}-bands were obtained using the Optical Robotic Observatory (ORO) and the \textit{V}-band data were taken from the ASAS database. We found that the system exhibits high levels of chromospheric emission, as is expected for binary systems. We found no evidence of eclipses in the photometry. Therefore, the inclination angle must be low ($i \leq 63$ degrees) and the masses of the primary and secondary components cannot be lower that 1.54 and 0.82 \textit{M}\textsubscript{\(\odot\)} respectively. These estimated masses agree with their spectral type.

\begin{table*}
\centering
\caption{Spectra analysis summarize. Column 1 is Heliocentric Julian date xJD = HJD - 2,450,000; column 2 and 3 are RV's and Ca II flux for primary component; column 4 and 5 are RV's and Ca II flux for secondary component.}
\begin{adjustbox}{max width=\textwidth}
\begin{tabular}{lcccc}
\hline\hline\noalign{\smallskip}
HJD - 2,450,000	& RV (primary) & $f$ (primary) & RV (secondary) & $f$ (secondary)\\
(days)	& (km\,s$^{-1}$) & (\(\num{e-12}\) erg cm\textsuperscript{-2} \AA\textsuperscript{-1} s\textsuperscript{-1}) & (km\,s$^{-1}$) &   \(\num{e-12}\) erg cm\textsuperscript{-2} \AA\textsuperscript{-1} s\textsuperscript{-1}) \\
\hline\noalign{\smallskip}
1769.9198	&	13.4	$\pm$	3.8	&	7.97	$\pm$	0.13	&	-50.9	$\pm$	5.5	&	7.97	$\pm$	0.10	\\
1972.5502	&	32.6	$\pm$	2.9	&	9.33	$\pm$	0.12	&	-81.5	$\pm$	5.2	&	9.36	$\pm$	0.16	\\
2363.5186	&	36.7	$\pm$	3.9	&	9.47	$\pm$	0.11	&	-87.3	$\pm$	6.1	&	8.11	$\pm$	0.13	\\
2519.8542	&	-36.2	$\pm$	3.0	&	9.46	$\pm$	0.23	&	39.2	$\pm$	6.1	&	8.39	$\pm$	0.19	\\
2600.7692	&	40.1	$\pm$	2.2	&	9.10	$\pm$	0.17	&	-104.1	$\pm$	6.2	&	8.68	$\pm$	0.18	\\
2714.5270	&	44.7	$\pm$	2.8	&	8.72	$\pm$	0.13	&	-101.3	$\pm$	5.7	&	8.43	$\pm$	0.13	\\
2897.7720	&	42.4	$\pm$	2.5	&	8.70	$\pm$	0.18	&	-103.0	$\pm$	6.3	&	8.01	$\pm$	0.14	\\
2980.7187	&	-6.5	$\pm$	3.8	&	8.36	$\pm$	0.09	&	-13.8	$\pm$	5.3	&	8.04	$\pm$	0.17	\\
3073.5522	&	-55.7	$\pm$	3.1	&	9.53	$\pm$	0.12	&	88.1	$\pm$	5.6	&	8.39	$\pm$	0.14	\\
3276.8205	&	-50.0	$\pm$	4.8	&	9.97	$\pm$	0.12	&	78.8	$\pm$	5.8	&	8.64	$\pm$	0.15	\\
3335.7444	&	-58.3	$\pm$	2.4	&	10.77	$\pm$	0.14	&	85.1	$\pm$	7.8	&	8.53	$\pm$	0.17	\\
3448.5788	&	-10.2	$\pm$	5.3	&	9.36	$\pm$	0.12	&	2.0	$\pm$	5.2	&	8.29	$\pm$	0.14	\\
3699.7699	&	41.4	$\pm$	4.5	&	9.24	$\pm$	0.12	&	-103.1	$\pm$	5.4	&	8.86	$\pm$	0.14	\\
4080.7643	&	42.8	$\pm$	4.5	&	10.17	$\pm$	0.11	&	-101.5	$\pm$	5.4	&	9.01	$\pm$	0.17	\\
4820.6945	&	-55.1	$\pm$	3.9	&	9.62	$\pm$	0.12	&	82.6	$\pm$	5.5	&	8.19	$\pm$	0.12	\\
4903.5406	&	10.6	$\pm$	5.4	&	10.06	$\pm$	0.13	&	-45.1	$\pm$	5.1	&	9.01	$\pm$	0.15	\\
5263.5212	&	36.5	$\pm$	4.4	&	10.26	$\pm$	0.15	&	-85.2	$\pm$	5.7	&	8.29	$\pm$	0.14	\\
5637.5522	&	-60.4	$\pm$	4.6	&	9.07	$\pm$	0.16	&	81.3	$\pm$	5.9	&	7.96	$\pm$	0.13	\\
6281.7795	&	-47.0	$\pm$	4.5	&	9.91	$\pm$	0.19	&	70.9	$\pm$	5.9	&	8.81	$\pm$	0.16	\\
6284.6758	&	6.7	$\pm$	4.2	&	9.62	$\pm$	0.14	&	-48.7	$\pm$	5.2	&	9.19	$\pm$	0.15	\\
6355.5412	&	-61.7	$\pm$	4.5	&	10.12	$\pm$	0.13	&	83.0	$\pm$	5.3	&	9.54	$\pm$	0.17	\\
6591.8426	&	-11.4	$\pm$	4.3	&	11.22	$\pm$	0.38	&	-15.0	$\pm$	4.5	&	8.96	$\pm$	0.21	\\
6735.5367	&	-10.8	$\pm$	4.6	&	8.55	$\pm$	0.14	&	-6.5	$\pm$	4.9	&	8.04	$\pm$	0.14	\\
7006.8318	&	34.8	$\pm$	3.7	&	10.77	$\pm$	0.18	&	-97.6	$\pm$	5.8	&	8.94	$\pm$	0.18	\\
8435.7348	&	-16.0	$\pm$	2.0	&	12.47	$\pm$	0.15	&	6.2	$\pm$	8.4	&	9.39	$\pm$	0.17	\\
8436.6167	&	32.9	$\pm$	2.6	&	13.16	$\pm$	0.26	&	-84.1	$\pm$	8.7	&	9.12	$\pm$	0.23	\\
\hline 
\end{tabular}
\end{adjustbox}
\centering
\label{resumen}
\end{table*}

We separated the spectra of both componentes, and we were able to determine accurate orbital parameters, in particular an orbital period of 4.95 days. The results are presented in Table \ref{resumen}. In the different photometry data sets, we observed a modulation with half the orbital period, due to the ellipticity of the stars.

Studying the GLS periodogram of the long-term $V$ magnitude, we found that the system exhibits a possible periodic behavior of 5.5 years (or \(\sim\) 2030 days). A cycle with a similar period can also be observed in the Ca \II\ fluxes of the secondary component, the least active star of the system. The presence of this modulation in the two independent data sets reinforces the detection of the cycle. 
On the other hand, there is a phase difference between the cycles in the photometry and the Ca \II\ flux, with the latter delayed by 554 days, about $P/4$. This timelag between photometric and magnetic variations has already been observed for stars of different spectral types \citep{Diaz2007}, although for cooler stars and with the opposite sense. It is also interesting that, although the Ca \II\ flux in both stars is fairly correlated, only one of the components shows a cycle, something which is also different from what we found for the cooler system studied in \citet{Diaz2007}. 

Therefore, the magnetospheres of the stars must be interacting. \citet{Vahia1995} simulated the interaction of the magnetic field of binary stars, modeled as perfect dipoles surrounded by vacuum. They showed that it should be quite common to find one of the components located completely inside the magnetosphere of the companion. \citet{Zaqarashvili2002} described the mechanisms by which magnetic activity can be enhanced in interacting systems like this one, which would explain the high emission levels in the Ca \II\ lines. A similar effect was observed by \citet{Diaz2007} in another binary system.

The activity period we found places the system in the active branch of the $P_\mathrm{rot}-P_\mathrm{cyc}$ diagram found by \citet{bohm2007} for solar-type stars, which is consistent with the level of activity of the stars under study.



\bibliographystyle{mnras}
\bibliography{mybiblio} 


\label{lastpage}
\end{document}